# Valley excitons in two-dimensional semiconductors

Hongyi Yu[1,2], Xiaodong Cui[1], Xiaodong Xu[3,4] and Wang Yao[1,2*]

[1]Department of Physics, The University of Hong Kong, Hong Kong, China
[2]Center of Theoretical and Computational Physics, The University of Hong Kong, Hong Kong, China
[3]Department of Physics, University of Washington, Seattle, Washington 98195, USA
[4]Department of Material Science and Engineering, University of Washington, Seattle, Washington 98195, USA

[*]Correspondence to: wangyao@hku.hk

**Abstract**: Monolayer group-VIB transition metal dichalcogenides have recently emerged as a new class of semiconductors in the two-dimensional limit. The attractive properties include: the visible range direct band gap ideal for exploring optoelectronic applications; the intriguing physics associated with spin and valley pseudospin of carriers which implies potentials for novel electronics based on these internal degrees of freedom; the exceptionally strong Coulomb interaction due to the two-dimensional geometry and the large effective masses. The physics of excitons, the bound states of electrons and holes, has been one of the most actively studied topics on these two-dimensional semiconductors, where the excitons exhibit remarkably new features due to the strong Coulomb binding, the valley degeneracy of the band edges, and the valley dependent optical selection rules for interband transitions. Here we give a brief overview of the experimental and theoretical findings on excitons in two-dimensional transition metal dichalcogenides, with focus on the novel properties associated with their valley degrees of freedom.

**Keywords**: exciton, valley physics, two-dimensional semiconductor, transition metal dichalcogenides

**Introduction**

Atomically thin group-VIB transition metal dichalcogenides (TMDs) have recently attracted vast interest as a new class of gapped semiconductors in the two-dimensional (2D) limit [1-3]. The compounds have the chemical composition of $MX_2$ where M stands for the metal atom W or Mo and X is S or Se. The stable phases of bulk crystals are of a layered structure where the elementary unit, the monolayer, is an X-M-X covalently bonded 2D hexagonal lattice. The monolayers are stacked and bounded by the weak van der Waals interaction. Monolayers can be extracted from bulk crystals by mechanical exfoliation [4-6], or synthesized using chemical vapor deposition or molecular beam epitaxy [7-12], similar to the preparation of graphene. When TMDs are thinned down from bulk to monolayers, a striking change in their electronic structure is the crossover from indirect to a direct band gap at the degenerate but inequivalent K and –K valleys at the corners of the hexagonal

Brillouin zone [4,5,8].

The direct band gap of monolayer TMDs is in the visible frequency range, ideal for the exploration of optoelectronic applications and semiconductor optics. Upon the absorption of a photon from an optical field, a valence band electron can be excited to the conduction band, and the vacancy left behind in the valence band is usually described as a hole which is a quasiparticle carrying positive charge. The attractive Coulomb interaction between the conduction band electron and the valence band hole can bound them into a hydrogen-like state, known as exciton, which is an elementary excitation that plays key role in optoelectronic phenomena. The bound electron-hole pair can also capture an extra electron or hole to form a negatively or positively charged exciton, also known as trion. Through the optical interband transition described above, excitons can be interconverted with photons. Neutral and charged excitons have been discovered in monolayer TMDs from the reflection and photoluminescence spectra [13-15], where the 2D geometry makes possible the remarkable electrostatic tunability between the neutral and the two charged configurations through gated controlled doping. The measured energy differences between the charged and neutral excitons point to an exceptionally large binding energy [13-15], which is also predicted by first-principles calculations [16,17], and jointly revealed by various measurements including the reflection spectra [18], two-photon absorption [19-22], and scanning tunneling spectroscopy [23,24]. The strong Coulomb binding arises from the reduced screening in the 2D geometry as well as the large effective masses of both the electron and the hole [17,20].

Another unique and interesting aspect of excitons in 2D TMDs is their valley configurations. As both the conduction and valence band edges are at the degenerate K and –K valleys, the lowest energy exciton states can be classified by the valley configurations as well as the spin configurations (Fig. 1). Those configurations with an electron and a hole in the same valley with opposite spin are *bright* excitons that can emit photon, as the spin and momentum conservation can be satisfied in the electron-hole recombination process. Interestingly, the interband optical transition in monolayer TMDs is associated with a valley selection rule originated from the three-fold rotational symmetry of the 2D lattice [25,26], such that the valley configurations of the bright excitons directly correspond to the circular polarization of the emitted/absorbed photon. This makes possible the optical addressability of the excitonic valley pseudospin [13,27-29], a property unique to 2D TMDs which have attracted remarkable interest.

Apart from the discovery of ultra-strong Coulomb binding and the optical addressability of valley pseudospin, other unique properties have been observed or predicted for valley excitons in 2D TMDs. These include the spin-layer locking effect and spectrally resolvable intra- and inter-layer trions in bilayer [30,31], the valley-orbital coupling and excitonic fine structure from the electron-hole exchange [32], the trion valley-Hall effect [32], anomalous Rydberg series of excitonic excited states [17-20,22,24], valley Zeeman splitting and magnetic tuning of polarization and coherence of the excitonic valley pseudospin [33-36], and valley selective optical

Stark effect [37,38].

Besides, functional optoelectronic devices based on valley excitons in 2D TMDs are being demonstrated. In lateral p-i-n junctions electrostatically formed in monolayer WSe$_2$, electroluminescence has been observed when the electrons and holes are injected into the intrinsic region from the n- and p-doped regions respectively under the forward bias [39-41]. Because of the strong Coulomb interaction, the electrons and holes form valley excitons before the radiative recombination [39-41]. A unique feature of these excitonic light-emitting p-n junctions is that the spectral weight of electroluminescence from the neutral and charged excitons is tunable by the bias current [39-41]. In some WSe$_2$ p-n junctions, the electroluminescence is found to be circularly polarized, where the polarization changes sign when the p-n junction is flipped [42-44]. According to the valley optical selection rule, the polarization implies that the emission from excitons in the two valleys is unbalanced and is electrically controllable, which realized a prototype valley light-emitting transistor [42-44]. Moreover, valley-selective second-harmonic generation at the ground-state exciton resonance has been demonstrated in monolayer WSe$_2$ field effect transistors. The second-harmonic signal generated from such devices can be tuned over an order of magnitude by electrostatic doping, and the tunability arises from the electrostatic charging effect which transfers excitonic oscillator strength between the neutral and the charged excitons [45].

The present article is motivated by these exciting physics and potential applications from the extraordinary properties of valley excitons in 2D TMDs. Below we give an overview of the theoretical understanding and experimental findings on the various aspects of valley excitons.

**The exciton spin and valley configurations**

In monolayer TMDs, the conduction band minima and valence band maxima are both located at the degenerate $K$ and $-K$ points at the corners of the hexagonal Brillouin zone. The $K$ and $-K$ valleys are time-reversal of each other. The conduction (valence) states at $\pm K$ valley mainly consist of transition metal $d_{z^2}$ ($d_{x^2-y^2} \pm i d_{xy}$) orbitals[46] with the magnetic quantum $m = 0$ ($m = \pm 2$). The strong spin-orbit-coupling (SOC) from the metal d-orbitals then leads to a large spin splitting for the valence band. The splitting value ranges from ~150 meV for MoX$_2$ to ~450 meV for WX$_2$ (X=S, Se). The mirror-symmetry in the out-of-plane (z) direction dictates that the splitting has to be in the z-direction [30,31], while the time-reversal symmetry (with inversion symmetry breaking) dictates that the spin splitting has opposite signs at valley K and -K (Fig. 1). For low energy configurations of excitons, because of the large spin splitting, we only need to consider the top spin sub-band for the valence states. For holes at such valence band edge, the spin and valley indices are then locked together, i.e. valley K (⁻K) only have spin up (down) states.

For the conduction states at the $\pm$K valleys, the dominant $d_{z^2}$ component do not contribute SOC, but the small components from the transition metal $d_{xz}$, $d_{yz}$ and the chalcogen $p_x$, $p_y$ orbitals give rise to a finite spin splitting [47,48]. The form of the SOC is the same as the valence band, but the magnitude is much smaller ($E_{c,\mathbf{K},\uparrow} - E_{c,\mathbf{K},\downarrow} \approx -3$ meV, $-21$ meV, $29$ meV and $36$ meV for monolayer $MoS_2$, $MoSe_2$, $WS_2$ and $WSe_2$ respectively from first principle calculations [47]). It should be noted that this conduction band spin splitting has an overall sign change between $MoX_2$ and $WX_2$. On the other hand SOC slightly renormalizes the effective mass of the two spin-split conduction bands, leading to conduction band crossings in $MoX_2$ but not for $WX_2$ [47] (the upper conduction band at $\pm K$ has larger/smaller effective mass than the lower one in $MoX_2/WX_2$, see Fig. 1). Both spin-split conduction bands are relevant for the low energy excitons.

Consider first the neutral exciton $X_0$. When the electron and hole constituents are at different valleys, their direct recombination is forbidden as momentum conservation can not be satisfied and the exciton is therefore dark (Fig. 1). When the electron and hole constituents are at the same valley with opposite (same) spin, the exciton is a bright (dark) one. Here we use the convention that a spin-down (-up) hole describes the absence of a spin-up (-down) valence electron. Optical interband transitions always conserve the spin as well as the valley index. Therefore, only the bright exciton can radiatively recombine and result in a photon emission, as the name itself implies. Bright excitons are directly observable both from the photoluminescence spectrum and the absorption (reflection) spectrum [4,5]. In addition to the lowest energy bright valley excitons (often referred as A excitons), the PL and absorption/reflection spectra also show the presence of a high-energy bright valley exciton configuration, i.e. B exciton, where the hole is in the higher-energy spin split valence band. The energy separation of the A and B excitons has been used to extract the valence band spin splitting [4,5,27,49]. The lifetime of B excitons is much shorter, as it will relax to the lower energy configurations through fast non-radiative channels. Below we will focus only on the A excitons that play more important roles in the optoelectronic phenomena. The creation operator for the $K$ valley bright exciton can be written as $\hat{B}^\dagger_{\mathbf{k},+} \equiv \sum_\mathbf{q} \psi_\mathbf{k}(\mathbf{q}) \hat{e}^\dagger_{\mathbf{K}+\mathbf{q}+\frac{\mathbf{k}}{2},\uparrow} \hat{h}^\dagger_{-\mathbf{K}-\mathbf{q}+\frac{\mathbf{k}}{2},\Downarrow}$ where $\mathbf{k}$ is the in-plane center-of-mass wave vector and $\psi_\mathbf{k}(\mathbf{q})$ describes the momentum space electron-hole relative motion, and $\hat{e}^\dagger_{\mathbf{K}+\mathbf{q}+\frac{\mathbf{k}}{2},\uparrow}$ ($\hat{h}^\dagger_{-\mathbf{K}-\mathbf{q}+\frac{\mathbf{k}}{2},\Downarrow}$) is the creation operator for an electron (hole) with the subscript denoting its momentum and spin. Considering the time reversal symmetry between the two valleys, the $-K$ valley bright $X_0$ is $\hat{B}^\dagger_{\mathbf{k},-} \equiv \sum_\mathbf{q} \psi^*_{-\mathbf{k}}(\mathbf{q}) \hat{e}^\dagger_{-\mathbf{K}-\mathbf{q}+\frac{\mathbf{k}}{2},\downarrow} \hat{h}^\dagger_{\mathbf{K}+\mathbf{q}+\frac{\mathbf{k}}{2},\Uparrow}$. The photon emission needs to satisfy both the energy and momentum conservation. Thus, even in the bright exciton branch, only those states that lie within the light cone (the conical region defined by $E > \hbar c |\mathbf{k}|$) can directly recombine and emit photons. For states outside the light cone, the radiative recombination needs the assistance of phonon scattering to satisfy the energy momentum conservation.

When the TMDs sample is negatively (positively) doped, $X_0$ can further bind an excess electron (hole) to lower its energy. The formed three-body bound quasi-particle is called negatively (positively) charged exciton or negative (positive) trion, denoted by X$^-$ (X$_+$). Just like bright exciton X$_0$, the bright trion is the one which can radiatively recombine, emitting a photon and leaving an electron/hole. In principle a bright trion with any center-of-mass wave vector can recombine since the wave vector can be transferred to the resulted electron/hole. But detailed analysis shows that the radiatively emission rate (or the trion brightness) exponentially decays with the increase of the center-of-mass wave vector [50], which gives rise to a low-energy tail in the trion PL spectrum [14].

In X$^-$ (X$_+$), to lower the energy, the two electrons (holes) shall have either opposite spin or valley index. Fig. 1 shows the various spin and valley configurations of trions. We only list those X$_0$ and X$^-$ with their hole component in the K valley, and those X$_+$ with electron in the K valley, while all other configurations are the time reversal counter parts of the ones shown. The energy splitting between the different spin and valley configurations ($\lambda_0$, $\lambda_\pm$ and $\lambda'_-$ in Fig. 1) is estimated to be in the order of a few tens meV, mainly due to the SOC splitting of the conduction bands, as well as the effective mass differences of the spin-split sub-bands. The latter effect comes in through the exciton binding energy. In MoX$_2$, the contribution to $\lambda_0$, $\lambda_\pm$ and $\lambda'_-$ from the binding energy differences is expected to have an opposite sign to the contribution from the conduction band spin splitting. The lack of accurate information on the exciton binding energies of the two spin sub-bands has resulted in the uncertainty in the magnitude and even sign of $\lambda$, which need to be determined in future experiments.

**Exciton binding energy**

The binding energy of a neutral or charged exciton characterizes how stable the bound state is, and is determined by the strength of the Coulomb interaction and the effective masses of the electron and hole. The binding energy $E_b$ of X$_0$ is defined as its lowered energy compared to the free electron-hole pair, while the trion binding energy (also called charging energy) $E_c$ is the difference between X$_\pm$ and the unbound state of an X$_0$ plus a free electron or hole. In the exciton luminescence, $E_c$ then corresponds to the spectra separation between the X$_0$ emission and the X$_\pm$ emission. Photoluminescence (PL) experiments clearly show well separated X$_0$ and X$_\pm$ emission peaks in 2D TMDs. From the energy splitting, the trion charging energy $E_c$ is determined to be 18 meV [15], 30 meV [14], 20-40 meV [51] and 30 meV [13] in monolayer MoS$_2$, MoSe$_2$, WS$_2$ and WSe$_2$ respectively. For the X$_0$ binding energy $E_b$, *ab initio* calculations give extremely large values in the order of ~0.5 to 1 eV [16,17,52-55]. The direct determination of $E_b$ from the PL spectra or absorption spectra has not been possible, as the edge of the band-to-band transition has not been unambiguously determined.

While the one-photon process measures the exciton ground state (1s) energy, in two-photon process, it is the exciton excited state (e.g. 2p) that becomes bright. This makes possible the determination of the energy separation between the 1s and 2p states, which provide a lower bound of several hundred meV on the $X_0$ binding energy $E_b$ [19-22]. Also one-photon reflectance contrast spectrum has been used to extract the exciton Rydberg series from 1s to 5s in monolayer WS$_2$ [18], a fitting then gives $E_g \sim 2.41$ eV and $E_b \sim 0.32$ eV.

Various methods have been used to extract the band-to-band transition energy, and its difference from the $X_0$ resonance measured by PL or absorption/reflection spectra then directly gives the exciton binding energy $E_b$. Two-photon-excitation induced PL measurements in monolayer WS$_2$ [21] and WSe$_2$ [19] have claimed the observation of features of band-to-band transition at the energies 2.73 eV and 2.02 eV which then lead to the binding energy $E_b \sim 0.71$ eV and 0.37 eV respectively. On the other hand, scanning tunneling microscopy/spectroscopy has been carried out to directly measure the quasiparticle bandgap (electronic bandgap) [23,24]. Scanning tunneling spectroscopy (STS) shows a 2.2 eV bandgap for monolayer MoSe$_2$ grown on bilayer graphene [24], resulting in $E_b \sim 0.55$ eV. Combined with *ab initio* calculation, $E_b$ is found to be ~0.65 eV without the screening of graphene [24]. Monolayer MoSe$_2$ grown on highly oriented pyrolytic graphite (HOPG) shows a similar quasiparticle band gap of 2.1eV in the STS measurement [12], which implies $E_b \sim 0.5$ eV. Monolayer MoS$_2$ grown on HOPG is shown to have a quasiparticle band gap of 2.15 eV or 2.35 eV depending on the threshold tunneling current [23,56], and the measured exciton PL peak is at 1.93 eV, which then leads to $E_b \sim 0.22$ eV or 0.42 eV. STS also shows an quasiparticle band gap of 2.51 eV (2.59 eV) at 77 K for monolayer WSe$_2$ (WS$_2$) on HOPG [56]. The exciton resonance in monolayer WSe$_2$ is measured to be ~1.65 eV at room temperature [56], which then leads to $E_b \sim 0.86$ eV.

These measured $X_0$ binding energy are in the same order of magnitude to those *ab initio* results. Such large values of $E_b$ (one order of magnitude stronger than in bulk TMDs [57] and conventional GaAs-type quantum wells) come from the large effective masses of both the electron and hole, the spatial confinement in the out-of-plane direction and the reduced screening of the dielectric environment. The 2D nature of monolayer TMDs enhances the binding energy to four times of that in 3D case. The dielectric mismatch between the TMDs and the substrates/air enhances $E_b$ even further, because the electric field line between the electron and the hole can penetrate into the air and the weakly screened substrates. The spatial-dependent effective dielectric constant (weaker at larger electron-hole separation) results in a significant deviation from a 2D hydrogen model, as indicated by the measured exciton Rydberg series [18,20,22]. The actual number on the exciton binding energy varies between measurements. The effects of substrates, sample fabrications on the exciton binding energy are not well understood. And in extraction of binding energy from the difference of exciton resonance from the quasiparticle bandgap measured in STS measurements, care shall be taken in the attribution of conduction and valence band edges in the STS as the higher energy critical points Γ and Q can contribute much larger weight in the scanning tunneling spectra than the K points [3,24]. These all need further experiments to clarify.

Usually excitons are classified into Frenkel excitons and Wannier-Mott excitons. In Frenkel excitons, the separation between the electron and hole is in the order of the unit cell and the binding energy is typically $\sim 0.1 - 1$ eV. For typical Wannier-Mott exciton, the electron-hole separation is much larger than lattice constant, and the binding energy is much weaker than that of Frenkel excitons. Here in 2D TMDs, the wavefunction is still largely of the Wannier-Mott type, while the binding energy is comparable to the typical Frenkel exciton. The calculated 2D Bohr radius is about $a_B \sim 1$ nm [17], and the wave function for the electron-hole relative motion extends over several tens unit cells [17,20]. For trions in TMDs, a variational method shows the average electron-hole distances are $\sim 1$ nm and $\sim 2.5$ nm [58].

**Exciton radiative and nonradiative decay**

Excitons can recombine radiatively and nonradiatively. The radiative lifetime of the bright $X_0$ inside the light cone is determined by its oscillator strength thus is also called intrinsic lifetime. A measurement using optical two-dimensional coherent spectroscopy gives an exciton homogeneous linewidth of ~meV in $WSe_2$ [59]. This homogeneous linewidth can be attributed to the overall effect from excitonic radiative, nonradiative decay and pure dephasing, it then gives a lower bound of ~0.2 ps to the exciton intrinsic radiative lifetime. There still lacks direct measurement on the intrinsic lifetime. For an ensemble of $X_0$, those with momentums outside the light cone are unable to radiatively recombine but these excitons can be scattered into the light cone, e.g. by impurity or phonon scattering, and thus they act as a reservoir, so the ensemble averaged radiative lifetime could be significantly increased [60]. The reported averaged radiative lifetime can reach $\sim$ns scale [61-64].

The nonradiative decay rates sensitively depend on experimental parameters like exciton density, temperature or sample quality. The observed time-resolved PL and absorption signals in monolayer TMDs exhibit multi-exponential decays [62,63,65,66], indicating complex exciton dynamics. An important intrinsic nonradiative decay channel is the exciton-exciton annihilation. When two excitons collide, one exciton can recombine and transfer the excess energy to the second exciton, which is then ionized and becomes a free electron-hole pair. The annihilation rate increases with ratio $E_b/E_g$ between the binding energy and the band gap, and is proportional to the exciton density. In monolayer TMDs $E_b/E_g$ is large because of the strong Coulomb binding, thus it is expected that this nonradiative process dominates when the exciton density is high. An exciton-exciton annihilation rate of $0.2 - 0.5$ cm$^2$/s has been estimated for $MoSe_2$ [66], $WS_2$ [21] and $WSe_2$ [62], which gives an exciton-exciton annihilation lifetime of $2 - 5$ ps for a density $\sim 10^{12}$ cm$^{-2}$. In monolayer $MoS_2$ the rate of exciton-exciton annihilation was reported to be $\sim 0.04$ cm$^2$/s [61]. Other nonradiative decay channels include interband carrier-phonon scattering[63], exciton captured by defect states [67,68] and relaxing to dark states [64,67].

**Bright exciton valley polarization and valley coherence**

For $X_0$, the bright A exciton has the valley degeneracy that can be described as a pseudospin **σ**. $\hat{\sigma}_z = +1$ and $\hat{\sigma}_z = -1$ correspond respectively to the exciton being in the K and –K valley (Fig. 2). The valley optical selection rule for inter-band transition correlates the excitonic valley pseudospin and the polarization of the photon: K (−K) valley bright $X_0$ can be interconverted with a $\sigma+$ ($\sigma-$) circularly polarized photon [25] (Fig. 2). This has been observed in polarization-resolved PL experiments in different monolayer TMDs [13,27-29,69].

The $X_0$ valley polarization is defined as $P_{+-} \equiv \frac{\langle \hat{B}^\dagger_{\mathbf{k},+} \hat{B}_{\mathbf{k},+}\rangle - \langle \hat{B}^\dagger_{\mathbf{k},-} \hat{B}_{\mathbf{k},-}\rangle}{\langle \hat{B}^\dagger_{\mathbf{k},+} \hat{B}_{\mathbf{k},+}\rangle + \langle \hat{B}^\dagger_{\mathbf{k},-} \hat{B}_{\mathbf{k},-}\rangle}$, which is directly reflected in the circular polarization of the luminescence. In the polarization resolved PL measurements, circularly polarized laser excites electron-hole pairs selectively in a valley, which then form valley polarized excitons. Inter-valley relaxation leads to the decay of the valley polarization $P_{+-}$. The observed large PL circular polarization indicates the valley relaxation time is larger than the exciton radiative lifetime [27,70-72]. However, as the exciton decay dynamics has not been observed, the direct extraction of valley relaxation timescale from the PL polarization may not be reliable.

A linearly polarized photon is in a coherent superposition of $\sigma+$ and $\sigma-$ polarization state. Thus, by the valley optical selection rule, linearly polarized optical field can generate $X_0$ in a coherent superposition of the two valley configurations, transferring the optical coherence to valley quantum coherence [13,25] (Fig. 2). The bright $X_0$ with an in-plane valley pseudospin $\hat{B}^\dagger_{\mathbf{k},V} \equiv \frac{e^{-i\theta}\hat{B}^\dagger_{\mathbf{k},+} + e^{i\theta}\hat{B}^\dagger_{\mathbf{k},-}}{\sqrt{2}}$ ($\hat{B}^\dagger_{\mathbf{k},H} \equiv \frac{e^{-i\theta}\hat{B}^\dagger_{\mathbf{k},+} - e^{i\theta}\hat{B}^\dagger_{\mathbf{k},-}}{\sqrt{2}}$) couples to the photon with linear polarization along $\hat{x}\cos\theta + \hat{y}\sin\theta$ ($\hat{x}\sin\theta - \hat{y}\cos\theta$). The PL linear polarization is then $P_{VH} \equiv \frac{\langle \hat{B}^\dagger_{\mathbf{k},V}\hat{B}_{\mathbf{k},V}\rangle - \langle \hat{B}^\dagger_{\mathbf{k},H}\hat{B}_{\mathbf{k},H}\rangle}{\langle \hat{B}^\dagger_{\mathbf{k},V}\hat{B}_{\mathbf{k},V}\rangle + \langle \hat{B}^\dagger_{\mathbf{k},H}\hat{B}_{\mathbf{k},H}\rangle} \leq \frac{2|\langle \hat{B}^\dagger_{\mathbf{k},+}\hat{B}_{\mathbf{k},-}\rangle|}{\langle \hat{B}^\dagger_{\mathbf{k},+}\hat{B}_{\mathbf{k},+}\rangle + \langle \hat{B}^\dagger_{\mathbf{k},-}\hat{B}_{\mathbf{k},-}\rangle}$. The maximum value of $P_{VH}$ as a function of $\theta$ direction gives the exciton valley coherence $|\langle \hat{B}^\dagger_{\mathbf{k},+}\hat{B}_{\mathbf{k},-}\rangle|$. Under the excitation by linearly polarized laser, it is observed that the PL of $X_0$ always carries the same linear polarization with the incident laser regardless of the crystalline orientation, an evidence of the optically generated valley coherence [13]. Since both the valley relaxation and valley pure dephasing lead to the decay of valley coherence, it is expected the valley pure dephasing time is also longer than the $X_0$ radiative lifetime.

In these polarization resolved PL measurements, the optically generated excitons will experience various scattering processes including carrier-carrier Coulomb interaction, carrier-phonon and carrier-impurities scattering. They are dominated by intra-valley scattering owing to the large **k**-space separation between the two valleys.

The intra-valley scattering doesn't couple to the valley pseudospin of excitons, thus preserving both the valley polarization and coherence. The inter-valley scattering induces both valley depolarization and valley pure dephasing, and it involves the short wavelength component of Coulomb potential ($V(\mathbf{K})$, which is typically very weak), or short wavelength phonons, or atomically sharp impurities. On the other hand, the electron-hole exchange interaction behaves like an in-plane effective magnetic field depending on the exciton center-of-mass motion, and gives rise to valley depolarization and decoherence [73]. Such effective magnetic field vanishes for $\mathbf{k} = 0$ excitons generated by perpendicularly incident lasers. Nevertheless, combined with momentum scattering it can still lead to the decay of PL circular and linear polarizations (see detail in the next section).

The bright trions also have the optical selection rule determined by its recombining electron-hole pair (Fig. 1). The observed trion PL shows strong circular polarization just like $X_0$. But the linear polarization for $X^-$ is absent owing to the exchange interaction induced energy splitting [13] (see details in the next section). $X_+$ can not emit linear polarized photon because it only has two valley configurations with the excess hole in opposite valleys. When a linear superposition of these two $X_+$ configuration emit a photon, the photon polarization is entangled with the valley pseudospin of the remaining hole, which eliminates the coherence between the $\sigma +$ and $\sigma -$ polarization states of the photon.

**Exciton fine structure from Coulomb exchange**

The binding of the electrons and holes into neutral and charged excitons is primarily due to the direct part of the Coulomb interaction. Coulomb interaction also has the exchange part. In the context of valley excitons, the exchange Coulomb interaction results in diagonal energy shift and off-diagonal coupling on the valley configurations, as summarized in Fig. 3 (a) and (b). There are 10 exchange-coupling terms, five are shown in the figure and the rest can be obtained by time reversal. In Fig. 3 (a) and (b) the right column are the diagrams for the Coulomb exchange between the conduction and valence electrons, and the left column schematically illustrate the corresponding effect on the electron and hole in their valley configurations. Processes II-IV are electron-hole exchange. Processes I and V correspond to electron-electron exchange, where process V is usually suppressed due to the energy mismatch from the conduction band spin splitting ranging from a few to a few tens of meV.

For the two bright $X_0$ in opposite valleys, only processes III and IV are involved (Fig. 3 (c)). Process III gives an overall energy shift independent of valley, while IV leads to a coupling between the two valleys. Using the valley pseudospin raising/lower operators $\hat{\sigma}_+ \equiv \hat{B}_{\mathbf{k}+}^\dagger \hat{B}_{\mathbf{k}-}$ and $\hat{\sigma}_- \equiv \hat{B}_{\mathbf{k}-}^\dagger \hat{B}_{\mathbf{k}+}$, the effect of electron-hole exchange interaction can be written as

$$\widehat{H}_{0,\text{ex}} = J_{\mathbf{k}}^{\text{intra}} + \left(J_{\mathbf{k}}^{\text{inter}}\hat{\sigma}_+ + \text{h.c.}\right) \tag{1}$$

Here $J_{\mathbf{k}}^{\text{inter}}$ ($J_{\mathbf{k}}^{\text{intra}}$) is the inter-valley coupling (intra-valley energy shift) which comes from the process IV (III). $\widehat{H}_{0,\text{ex}}$ splits the bright $X_0$ into two branches. Each branch is the equal superposition of the two valleys, i.e. the valley pseudospin lies in the plane ($\langle\hat{\sigma}_z\rangle = 0$), and has a dispersion $\hbar\omega_0 + \frac{\hbar^2 k^2}{2M_0} + J_{\mathbf{k}}^{\text{intra}} \pm |J_{\mathbf{k}}^{\text{inter}}|$ (Fig. 4 (a) and (b)).

The exact forms of $J_{\mathbf{k}}^{\text{inter}}$ and $J_{\mathbf{k}}^{\text{intra}}$ depend on the $X_0$ wave function. Symmetry analysis shows that up to the leading order of $k$, $J_{\mathbf{k}}^{\text{inter}}$ has a form of $J_{\mathbf{k}}^{\text{inter}} \propto -|\psi(\mathbf{r}_{\text{eh}} = 0)|^2 V(\mathbf{k}) k^2 e^{-2i\theta}$ with $\psi(\mathbf{r}_{\text{eh}})$ the real space wave function of the electron-hole relative motion, $V(\mathbf{k})$ the $\mathbf{k}$-space Coulomb potential, $\theta \equiv \text{atan}\frac{k_x}{k_y}$, and $J_{\mathbf{k}}^{\text{intra}} = |J_{\mathbf{k}}^{\text{inter}}| + \text{Const.}$ [32]. The factor $e^{-2i\theta}$ is from the requirement of in-plane rotational symmetry, which further dictates that each split $X_0$ branch has a chirality index two. Inside the light cone ($k \lesssim \omega_0/c \sim 10^{-3}K$), the upper (lower) branch is coupled to linear polarized photons with polarization directions longitudinal (transverse) to the $\mathbf{k}$ direction, and the splitting $2|J_{\mathbf{k}}^{\text{inter}}|$ corresponds to the longitudinal-transverse (LT) splitting, which is well known in GaAs quantum wells. In monolayer TMDs, because of the large Coulomb interaction the LT splitting is greatly enhanced, about two orders of magnitude larger than in GaAs quantum wells [32].

For the unscreened Coulomb interaction of form $V(\mathbf{k}) = \frac{2\pi e^2}{\epsilon k}$, it is found $J_{\mathbf{k}}^{\text{inter}} = -\frac{k}{K} J e^{-2i\theta}$ with $J \sim 1$ eV and $K$ the distance between the $\Gamma$ and $\pm K$ points of the Brillouin zone [32]. The dispersion is shown in Fig. 4 (a) and (b). Inside the light cone the upper branch shows a close to linear dispersion, thus can be viewed as a massless Dirac particle with chirality two. The LT splitting near the edge of the light cone is estimated to be $\sim$ meV, which could be much larger than the bright $X_0$ intrinsic line width.

The LT splitting can induce valley depolarization and decoherence [73-75]. Depending on the momentum scattering rate $\tau^{-1}$, the system can be divided into strong scattering ($\bar{\Omega}\tau \ll 1$) or weak scattering ($\bar{\Omega}\tau \gtrsim 1$) regime, where $\bar{\Omega}$ is the ensemble averaged LT splitting value. In the strong scattering regime, the $X_0$ dynamics can be described by the motional narrowing effect and the valley relaxation rate is given by $\sim \bar{\Omega}^2 \tau$ [73-75]. Since in monolayer TMDs the splitting is very large, it is possibly in the weak scattering regime. Further theoretical studies are needed to understand the role of LT splitting on the exciton valley relaxation and decoherence in the weak scattering regime.

The three-fold rotational (C3) symmetry of the lattice dictates that $J_{\mathbf{k}}^{\text{inter}}$ vanishes at $\mathbf{k} = 0$, as $X_0$ with zero center-of-mass wave vector inherits the C3 symmetry of $\pm K$ points. A finite $J_{\mathbf{k}=0}^{\text{inter}}$ would lead to bright $X_0$ eigenstates which can emit linearly polarized photons thus violates the rotational symmetry. An in-plane uniaxial strain breaks the rotational symmetry and gives rise to a nonzero $J_{\mathbf{k}=0}^{\text{inter}}$, which acts like an in-plane Zeeman field on the valley pseudospin and modifies the bright $X_0$ dispersion [32] (see Fig. 4 (c) and (d)).

Exchange interaction also affects trions. For the low energy configurations of $X_+$, there is only the diagonal energy shift: process II for dark $X_+$ and process III for bright $X_+$. They induce different energy shifts for bright and dark $X_+$. $X^-$ has more valley and spin configurations. The bright $X^-$ configurations are formed when a bright $X_0$ with valley pseudospin $\hat{\sigma}_z$ binds a low energy excess electron with spin $\hat{s}_z$, thus they can be characterized by these two indices. Here we focus on the ground state configurations of bright $X^-$. In Mo$X_2$, the ground state of bright $X^-$ has two-fold degeneracy ($\hat{\sigma}_z \hat{s}_z = -1$, c.f. Fig. 1), similar to bright $X_+$, and the exchange interaction only leads to a trivial energy shift. W$X_2$ is different because of the opposite conduction band spin-splitting from the Mo$X_2$ case (Fig. 1), and the ground states of bright $X^-$ can have four configurations. First, the exchange processes I and II exist only for the configurations with $\hat{\sigma}_z \hat{s}_z = 1$, which leads to a energy splitting $\delta$ between $\hat{\sigma}_z \hat{s}_z = 1$ and $-1$ (Fig. 3 (c)). Second, process IV couples configurations $\hat{\sigma}_z = 1$ and $-1$ with the same $\hat{s}_z$. Third, process III induces a global energy shift independent of $\hat{\sigma}_z$ and $\hat{s}_z$. For bright $X^-$ with a center-of-mass wave vector $-\hat{s}_z \mathbf{K} + \mathbf{k}$, its total exchange Hamiltonian is

$$\hat{H}_{-,\text{ex}} = J_{\mathbf{k}}^{\text{intra}} + \frac{\delta}{2}(\hat{\sigma}_z \hat{s}_z + 1) + \left(J_{\mathbf{k}}^{\text{inter}} \hat{\sigma}_+ + \text{h.c.}\right). \qquad (2)$$

The splitting $\delta$ is nearly independent on $\mathbf{k}$ because processes I and II correspond to Coulomb scattering with a large wave vector $\sim K$. Its strength is estimated to be $\delta \sim 6$ meV [32]. The values of $J_{\mathbf{k}}^{\text{inter}}$ and $J_{\mathbf{k}}^{\text{intra}}$ depend on the X-wave function. Nevertheless from symmetry analysis, $J_{\mathbf{k}=0}^{\text{inter}} = 0$ because of the three-fold rotational symmetry, and $J_{\mathbf{k}}^{\text{inter}} \approx -\frac{k}{K} J e^{-2i\theta}$ with comparable magnitude to that of X0. And $J_{\mathbf{k}}^{\text{intra}} \approx \left|J_{\mathbf{k}}^{\text{inter}}\right| + \text{Const.}$.

Processes I and II between the recombining electron-hole pair and the excess electron therefore act as an out-of-plane Zeeman field (with the sign dependent on $\hat{s}_z$) on the valley pseudospin $\hat{\sigma}_z$. The coherent superposition of the two configurations with $\hat{\sigma}_z = 1$ and $-1$ with the same $\hat{s}_z$ is then destroyed by this effective Zeeman field. Therefore, X- PL can not be linearly polarized in monolayer TMDs [13].

The direct observation of the excitonic fine structure is still challenging with the existing sample qualities. The estimated energy splitting is $\sim$meV while the measured PL linewidth is at least $\sim 5$ meV in MoSe$_2$ and WSe$_2$ [13,14] and much larger in MoS$_2$ and WS$_2$ [15,21]. Nevertheless, there have been indirect evidences supporting these predictions. As mention above, the splitting between X⁻ configurations due to the exchange interaction (processes I and II, c.f. Figure 3) explains the absence of linear polarization in the X⁻ emission in contrast to the $X_0$ emission [13]. It is also consistent with the observation that in bilayer WSe$_2$ the interlayer trion can have linear polarization, because such exchange interaction is suppressed when the recombining electron-hole pair and the excess electron are located in opposite layers [31]. Moreover, in a perpendicular magnetic field, the polarizations of the $X_0$ and X⁻ PL peaks show distinct magnetic field dependence, which can be explained by the different exchange fine structures of $X_0$ and X⁻ (see the section entitled "Magneto response of valley excitons") [33].

**Berry curvature and valley Hall effect of trion**

In an in-plane electric field **E**, a charged particle can acquire a transverse anomalous velocity due to the Berry phase effect, in addition to the normal velocity in the longitudinal direction. This may be described by the semiclassical equation of motion for a wavepacket: $\hbar \dot{\mathbf{r}} = \frac{\partial E_\mathbf{k}}{\partial \mathbf{k}} - e\mathbf{E} \times \mathbf{\Omega}_\mathbf{k}$ [76], where $E_\mathbf{k}$ is the energy dispersion, $e$ is the charge of the carrier (negative for electron), and **r** and **k** are the central coordinates of the wavepacket in real and momentum space respectively. $\mathbf{\Omega}_\mathbf{k}$ is the Berry curvature that characterizes the Berry phase effect [76]. It arises from the dependence of the internal structure of particle wavefunction on the center of mass wavevector. For Bloch electron, $\mathbf{\Omega}_\mathbf{k} = i\langle \nabla_\mathbf{k} u_\mathbf{k}| \times |\nabla_\mathbf{k} u_\mathbf{k}\rangle$, where $u_\mathbf{k}$ is the periodic part of the Bloch wave function. In monolayer TMDs, because of the inversion symmetry breaking, the conduction and valence band both acquire finite Berry curvatures at the K and –K valleys [2], where the time reversal symmetry dictates that the curvature must have opposite signs at K and –K in each band. Thus, in an in-plane electric field, carriers at the two valleys will flow in opposite transverse directions, i.e. a valley Hall effect [2,25,26,77,78].

Trion is also a charged particle. The dependence of the internal structure of trion wavefunction on the center-of-mass wavevector can give rise to a similar gauge structure to that of the electron [79]. An in-plane electric field can induce an anomalous transverse motion and give rise to the Hall effects of trion. In monolayer TMDs, the trions as composite particles can acquire valley dependent Berry curvature from two distinct origins, from the inheritance of the Berry curvature from the Bloch band, and from the Coulomb exchange interaction between the electron and hole constituents.

For the positively charged trion X$_+$, its Berry curvature is mainly inherited from the Bloch bands [79]. The electrons and holes in ±K valley have nearly **k**-independent values of Berry curvature $\Omega_{\pm\mathbf{K}+\mathbf{k}} \sim \pm 10 \text{ Å}^2$ [25]. The bright X$_+$ then acquires a Berry curvature given by the sum of the curvatures of its electron and hole constituents. As the Berry curvatures from the two holes in opposite valleys cancel, the $X_+$ Berry curvature is determined by the electron constituent, and thus the two valley configurations of bright $X_+$ have opposite Berry curvatures, giving rise to the valley Hall effect. The trion valley Hall effect can lead to valley polarization of trions with opposite signs at the two edges, which can be detected as the circularly polarized luminescence since the trion is associated with the valley optical selection rule. Also the excess hole left behind upon the trion recombination is valley and spin polarized, and the trion valley Hall effect can therefore be exploited for the generation of spin and valley polarization of carriers.

Similarly bright $X^-$ will also inherit the Berry curvature from its electron and hole constituents. Moreover, in $WX_2$, a much stronger Berry curvature can arise from the exchange interactions between the electrons and holes as shown in Eq. (2). For bright $X^-$ with a center-of-mass wave vector $-\hat{s}_z\mathbf{K} + \mathbf{k}$, the out-of-plane Zeeman exchange term $\frac{\delta}{2}\hat{\sigma}_z\hat{s}_z$ together with the inter-valley exchange term $J_{\mathbf{k}}^{\text{inter}}$ give rise to a Berry curvature [32]

$$\Omega_{-\hat{s}_z\mathbf{K}+\mathbf{k}} = -\hat{\sigma}_z \frac{2J^2}{K^2\delta^2}\left(1 + \frac{4k^2J^2}{K^2\delta^2}\right)^{-\frac{3}{2}}. \tag{3}$$

The sign of $\Omega_{-\hat{s}_z\mathbf{K}+\mathbf{k}}$ is determined by $\hat{\sigma}_z$, the valley index of the recombining electron-hole pair (Fig. 5), and it is independent of spin index $\hat{s}_z$ of the excess electron. With the estimated exchange coupling strength $J \sim 1$ eV and $\delta \sim 6$ meV, the peak value of $|\Omega_{-\hat{s}_z\mathbf{K}+\mathbf{k}}|$ can be $\sim 10^4$ Å$^2$ in the neighborhood of $\mathbf{k} = 0$, which is several orders larger than the curvature of the electron or hole. The luminescence polarization of bright $X^-$ is determined by $\hat{\sigma}_z$ through the valley optical selection rule, thus the $X^-$ valley Hall effect may also be detected from a spatially dependent PL polarization pattern (Fig. 5), similar to the $X_+$ valley Hall effect.

Various relaxation mechanisms can inhibit the observation of the trion valley Hall effect. If the population decay is fast, the trions with opposite Berry curvatures can not move sufficiently away from each other during their lifetime. A valley flip process (valley relaxation) changes the sign of the Berry curvature and the direction of the transverse motion, suppressing the valley Hall current. Also we note that the upper and lower energy branches of $X^-$ dispersions have the opposite Berry curvature, and the two branches are separated only by a small energy gap of a few meV in the light cone. Non-adiabatic dynamics can then cause the transitions between the two branches, which will also diminish the valley Hall effect. The non-adiabaticity can come from the momentum scattering which changes the $X^-$ wave vector, so the scattering rate shall not be too large in order to observe the trion valley Hall effect. A room temperature $X_0$ diffusion coefficient of $D \sim 12 - 15$ cm$^2$/s has been measured in monolayer MoSe$_2$ [80] and WSe$_2$ [81], the relation $D = \tau k_B T/M_0$ then leads to the $X_0$ scattering time $\tau \sim 0.2$ ps. Unlike $X_0$, $X_\pm$ are charged particles so the interactions with charged impurities and piezoelectric types of phonons should be stronger. We expect the $X_\pm$ momentum scattering time to be shorter than $X_0$. A low temperature and clean sample shall facilitate the observation of trion valley Hall effect.

**Magneto response of valley excitons**

In this section, we discuss the response of valley excitons to external magnetic fields. Compared to excitons in conventional semiconductors such as GaAs, a remarkably new feature in monolayer TMDs is that an exciton configuration and its time reversal counterpart have different valley configurations. The momentum mismatch between these distinct valley configurations will suppress their off-diagonal coupling in various occasions. For example, in monolayer TMDs, the two valley configurations of bright $X_0$ cannot be coupled by the in-plane magnetic field as it cannot supply the momentum mismatch. Experimentally, it has been shown that the exciton valley polarization in monolayer $MoS_2$ has no response to in-plane magnetic field up to 9 T [28,69].

On the other hand, the valley excitons show interesting response to out-of-plane magnetic field. In monolayer TMDs, the mirror-symmetry about the metal atom plane dictates that the magnetic moments of the carriers are in the out-of-plane (z) direction. And the time reversal symmetry requires the magnetic moments of the $\pm K$ valleys to have the same magnitude but opposite signs. An out-of-plane magnetic field then lifts the valley degeneracy, i.e. a valley Zeeman effect. With the valley optical selection rule, the valley Zeeman effect may be detected from the polarization resolved PL measurement, where the magnetic field is expected to split the $\sigma+$ and $\sigma-$ PL peaks that correspond to the interband transition energies in valley K and –K respectively [33-36].

The overall valley Zeeman splitting has three contributions [33]. The first is the spin magnetic moment. This contribution does not affect the optical resonances, because optical transitions conserve spin so that the shift of the initial and final states due to the spin magnetic moment is the same [25]. The second is the magnetic moment of atomic orbital or the intra-cellular component [33-36]. The conduction (valence) band in $\pm K$ valley mainly consists of transition metal $d_{z^2}$ ($d_{x^2-y^2} \pm id_{xy}$) orbitals with the magnetic quantum $m = 0$ ($m = \pm 2$). This contributes to a Zeeman shift of 0 and $\pm 2\mu_B B$ for the conduction and valence band, respectively. Therefore it is a major contribution to the magneto-splitting of the $\sigma+$ and $\sigma-$ PL peaks. The third is the valley magnetic moment associated with the Berry phase effect (or the lattice contribution) [33-36]. Note that within the minimum two-band **k·p** model for band edge electrons and holes in monolayer TMDs (i.e. massive Dirac fermion model [25]), because of the particle-hole symmetry, the valley magnetic moment (lattice contribution) contributes identical Zeeman shift for conduction and valence band. Nevertheless, corrections beyond this model result in a finite difference for the electron and hole valley magnetic moment, and it is this difference, as well as the atomic orbital contribution, that is measured by the splitting of the $\sigma+$ and $\sigma-$ PL peaks [33-36].

It is also found that such a perpendicular magnetic field can be used for tuning the polarization and coherence of the excitonic valley pseudospin. Monolayer MoSe$_2$ shows no PL polarization in the absence of the magnetic field. Applying a perpendicular magnetic field lifts the valley degeneracy and creates a population imbalance in the two valleys, which then leads to a field dependent PL polarization [35,36]. In monolayer WSe$_2$ the valley polarization is preserved during the exciton and trion lifetime, so the PL polarization sign is determined by the helicity of the excitation laser. The polarization of $X_0$ either increase or decrease with magnetic field depending on the relationship between the pump light helicity and the magnetic field direction [33]. In contrast, the $X^-$ PL polarization always increases with magnetic field [33,34]. Such distinct behaviors of bright $X_0$ and $X^-$ are explained by their qualitative different dispersion relations induced by the exchange interaction (see Eq. (1) and (2) in the previous section), which result in different valley depolarization processes [33]. On the other hand the valley coherence is always suppressed when applying a out-of-plane magnetic field, because the lifting of the valley degeneracy destroys the coherent pathways in the $X_0$ formation process [33].

**Excitons in bilayer TMDs homo- and hetero-structures**

When monolayer TMDs are stacked on top of each other, the van der Waals interaction can bound them into homostructures as well as various heterostructures. These offer new possibility to study the physics of excitons, in particular the interlayer excitons where the composite particles reside in different layers. Interlayer trions have been observed in WSe$_2$ homo-bilayers [31], and neutral interlayer exciton has been observed in TMDs hetero-bilayers [82-86]. These interlayer excitons and trions exhibit remarkable properties including the tunability by the interlayer bias. Below, we briefly introduce these interlayer excitons.

In TMDs homo- or hetero-bilayers, the interlayer coupling between states at the K-valleys is weak compared to the energy mismatch and hence the hybridization between the layers are negligible for these states [3]. Concerning the valley excitons with the electron and hole all from the K-valleys of the two layers, each of these electron or hole constituents is largely localized either in the upper or lower layer. The layer separation (~7 Å) is comparable to the intralayer exciton Bohr radius (~1 nm), thus the strong Coulomb interaction between the electrons and holes in different layers can bind them into interlayer neutral or charged excitons.

The interlayer $X_0$ has been observed in monolayer MoSe$_2$-WSe$_2$ vertical heterostructures through PL and PL excitation spectroscopy [82-86]. These heterostructures have the type-II band alignment, therefore the lowest energy configuration of $X_0$ is an interlayer one, with the electron and hole residing in opposite layers. Taking the MoSe$_2$-WSe$_2$ vertical heterostructures for example [82], the conduction band minimum (valence band maximum) is located in the MoSe$_2$ (WSe$_2$) layer, thus the interlayer $X_0$ has an energy much lower than the intralayer $X_0$. While interlayer hopping is substantially quenched for both the electron and hole by the conduction band offset and valence band offset respectively, its residue effect leads to small layer hybridization that allows the direct radiative recombination of interlayer $X_0$. The radiative recombination of interlayer $X_0$ gives a distinct PL peak. The spatially indirect nature reduces the optical dipole of interlayer $X_0$, thus substantially extends its lifetime, which is observed to exceed nanosecond [82]. Another unique aspect of the interlayer $X_0$ is that it corresponds to a permanent electric dipole in the out-of-plane direction. This makes possible the tuning of its energy by the interlayer bias. Moreover, it also gives rise to strong dipole-dipole repulsive interaction, observed as a blue shift in the exciton resonance with increasing power [82]. The repulsive interaction and the ultralong lifetime of interlayer exciton makes it an ideal candidate to explore the exotic phenomenon of excitonic Bose Einstein condensate [87].

In exfoliated $WSe_2$ homo-bilayer, interlayer trion has been observed [31]. TMDs bilayers exfoliated from the natural crystals are mostly 2H stacking, where the two layers are 180 degree rotation of each other. This 180 degree rotation switches the K and –K valleys. With the spin-valley coupling in each monolayer, the spin-splitting in 2H bilayer has a sign depending on both the valley and the layer index, which quenches the interlayer hopping at the K-valleys and results in the spin-layer locking effect: the spin up and down states in each valley are localized in opposite layers [2,3]. An out-of-plane electric field can then induces a spin Zeeman splitting [30]. The electrically induced Zeeman splitting of the conduction band $\Delta_{Ec}$ is larger than the valence band one $\Delta_{Ev}$ owing to the fact that the electrons (holes) have a close to zero (weak but finite) interlayer hybridization.

Considering the intralayer $X_0$, the one localized in the upper layer has an energy difference $\Delta_{Ec} - \Delta_{Ev}$ from that in the lower layer in presence of the out-of-plane electric field. In doped bilayer, the intralayer $X_0$ can bind a low energy excess electron or hole in the same layer (in the opposite layer) to form an intralayer (interlayer) trion. The interlayer trion has a smaller charging energy $E_{c,inter}$ than the intralayer trion $E_{c,intra}$. The total energy difference between the inter- and intra-layer trion is $(\Delta_{Ec} - \Delta_{Ev}) + (E_{c,intra} - E_{c,inter})$ for $X^-$, and $(\Delta_{Ec} - \Delta_{Ev}) - (E_{c,intra} - E_{c,inter})$ for $X_+$. The energy difference is larger for $X^-$ than $X_+$. In bilayer $WSe_2$, the splitting between the interlayer and interlayer $X^-$ have been observed in the PL, where the $X^-$ peak splits into a doublet with the increase of interlayer bias [31].

Large valley polarizations are observed for both the interlayer and the intralayer $X^-$ under circularly polarized pump [31,88]. Under linearly polarized pump, however, valley coherence is observed only for the interlayer $X^-$, but not for the intralayer one [31]. This is because the intralayer $X^-$ with all three particles localized in the same layer is similar to the trion in a monolayer, where the exchange coupling with the excess electron destroys the valley coherence of the recombining electron-hole pair. In contrast, for interlayer $X^-$, with the excess electron in the opposite layer, its exchange coupling with the recombining electron-hole pair is suppressed and the valley coherence can therefore be preserved.

*In summary*, we provide here an overview of the properties of valley excitons in 2D group-VIB TMDs. For this emerging topic, materials issues and the lack of physical insights on many observed properties make a comprehensive review of various aspects of excitons in these new materials impractical at this stage. Instead, we have focused on the physics associated with valley degrees of freedom, which distinguish 2D TMDs excitons from those in conventional semiconductors. A lot of issues need to be addressed by future experimental and theoretical studies. For example, one of the most urgent issues that need to be thoroughly studied is the exciton relaxation and decoherence mechanisms. The exciton relaxation dynamics in monolayer TMDs are shown to be complicated, and the sample quality and the excitation power both affect the exciton decay timescales. Multiple timescales are observed in time-resolved spectroscopy, showing the relevance of non-radiative decay channels. Further experiments are needed for understanding the various mechanisms of exciton non-radiative decay, as well as the possible roles of dark excitons in the radiative decay. The observation of valley polarization and coherence of excitons in photoluminescence experiments imply that the $T_1$ and $T_2$ times of excitonic valley pseudospin are slow compared to the exciton radiative and non-radiative decay. Polarization resolved pump-probe and spectral hole burning measurements are providing useful information on the excitonic valley dynamics [64,71,89,90]. But a thorough understanding of the mechanisms and timescales for the relaxation and decoherence of excitonic valley pseudospin is still lacking, which is key to the exploration of the applications and new physics associated with valley excitons.

**Funding:** HY and WY acknowledge support by the Croucher Foundation (Croucher Innovation Award), the University of Hong Kong (OYRA), and the RGC of Hong Kong (HKU17305914P). XC acknowledges support by RGC (HKU9/CRF/13G ) and UGC (AoE/P-04/08) of Hong Kong. XX acknowledges support from DoE, BES, Materials Science and Engineering Division (DE-SC0008145), the National Science Foundation (DMR-1150719), and the National Science Foundation, Office of Emerging Frontiers in Research and Innovation (EFRI – 1433496).

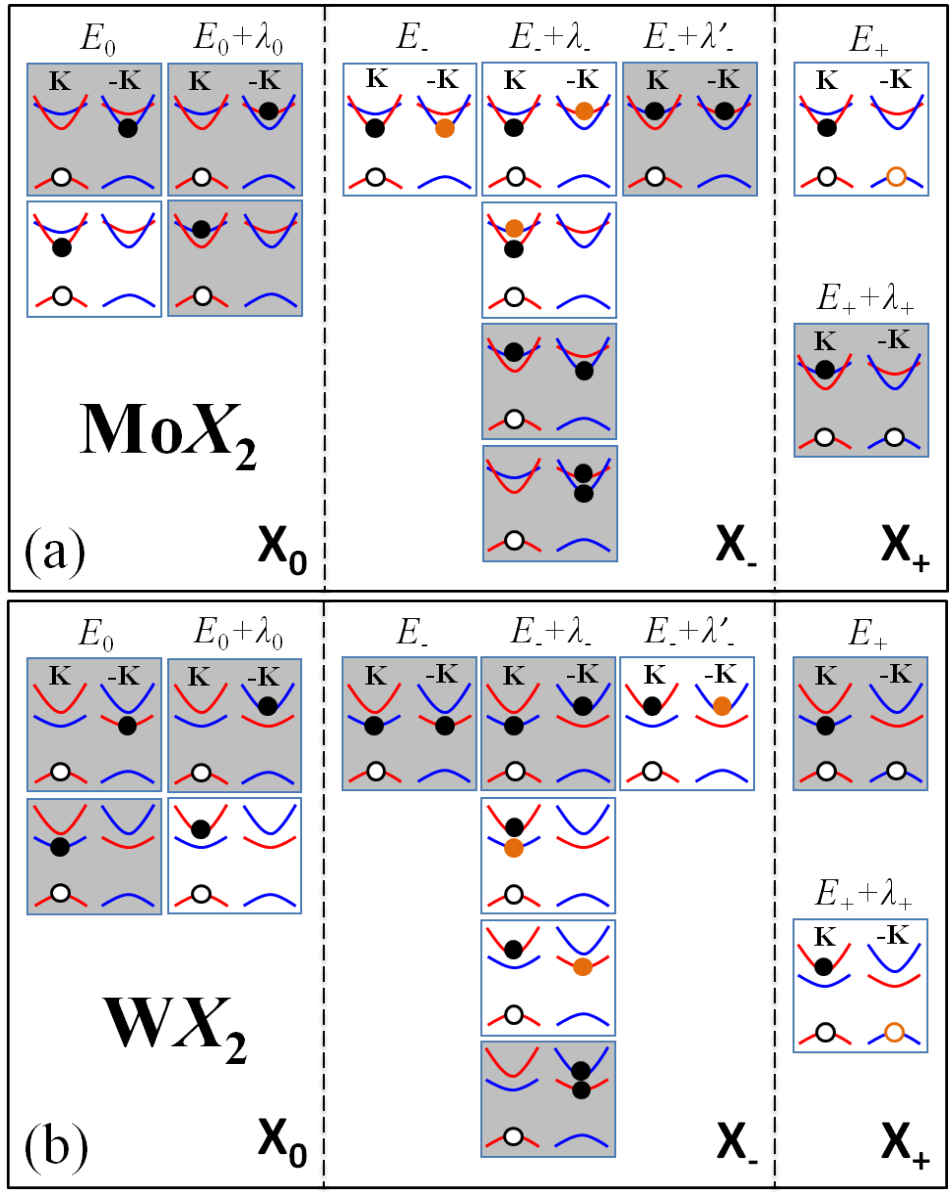

Fig. 1 | Valley configurations of neutral ($X_0$), negatively charged ($X^-$) and positively charged ($X_+$) excitons in monolayer Mo$X_2$ (a) and W$X_2$ (b) ($X$=S or Se). Solid (empty) dots denote electrons (holes), and red (blue) curves denote spin-up (spin-down) conduction and valence bands. For $X_0$ and $X_-$ only those with the hole in the K valley are shown, and for $X_+$ only those with the electron in the K valley are shown. The remaining configurations are the time reversal of those shown in the figure. Valley configurations with white (grey) background are bright (dark) excitons. In the bright $X_\pm$ the electron-hole pair that can recombine is shown in black color while the excess electron or hole is shown in orange color. Note that the two spin-split conduction bands have different effective mass due to the renormalization of the spin-orbit-coupling, leading to conduction band crossings in Mo$X_2$ but not for W$X_2$.

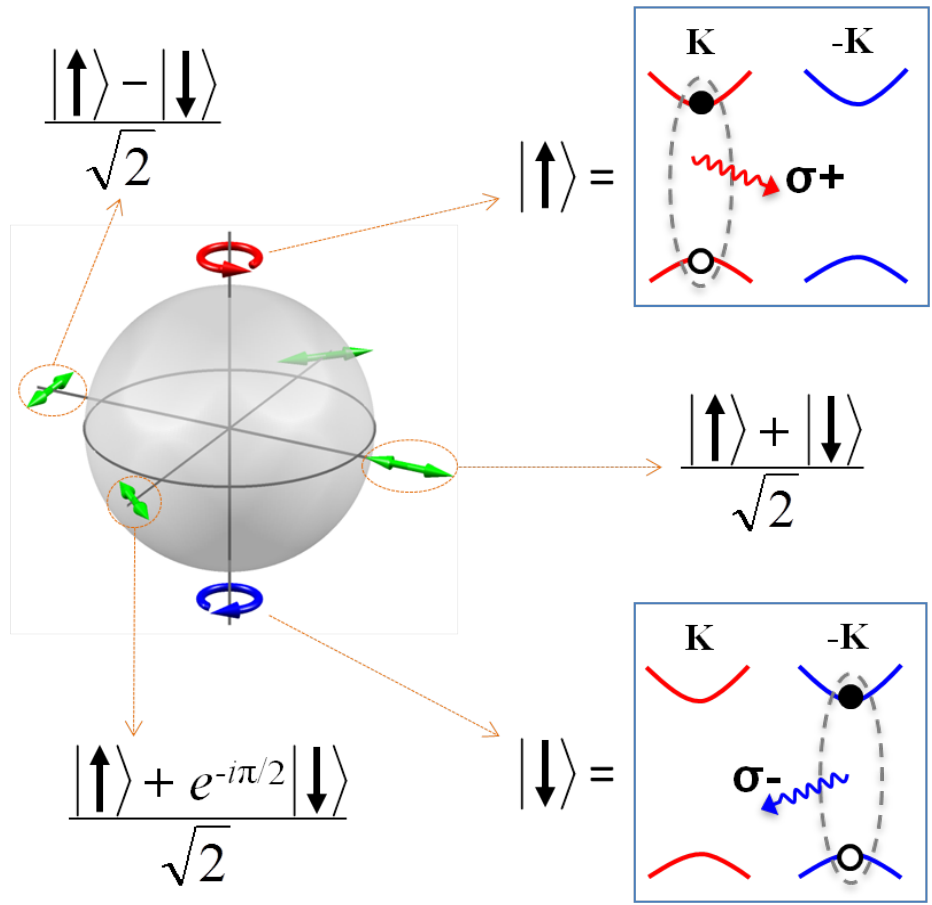

Fig. 2 | Optical addressability of valley pseudospin of bright $X_0$. A bright $X_0$ in K valley corresponds to valley pseudospin up ($\hat{\sigma}_z = +1$, north pole on the Bloch sphere), which emits $\sigma+$ circularly polarized photon, while bright $X_0$ in -K valley corresponds to valley pseudospin down ($\hat{\sigma}_z = -1$, south pole on the Bloch sphere), which emits $\sigma-$ circular polarized photon. The Bloch sphere equator corresponds to bright $X_0$ with an in-plane valley pseudospin (equal superposition of the two valleys), which emits linearly polarized photon with the polarization indicated by green double arrows.

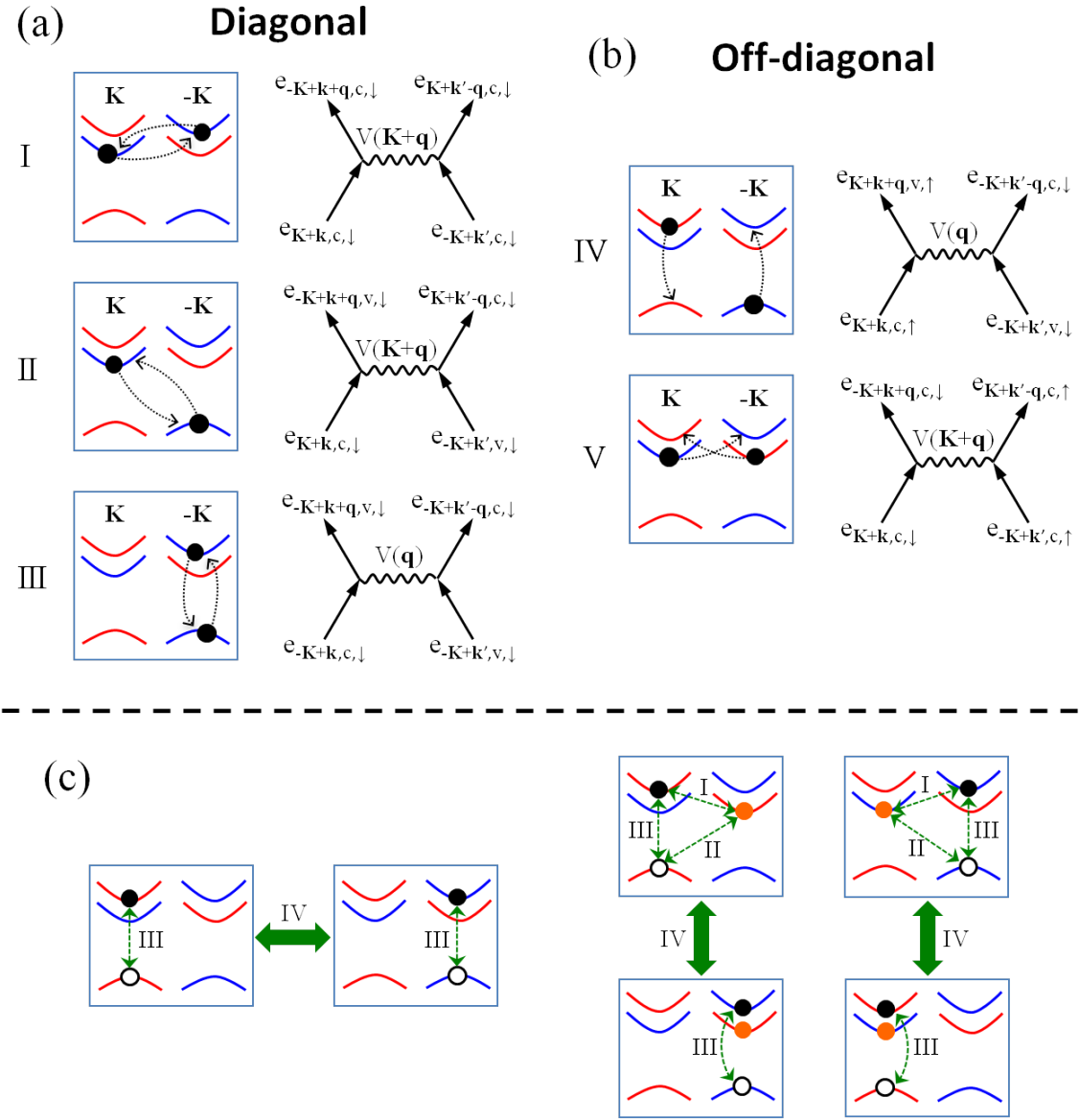

Fig. 3 | The diagonal energy shift (a) and off-diagonal coupling (b) due to the Coulomb exchange interaction between electron and hole and between electrons. (c) The effect of Coulomb exchange on $X_0$ and $X_-$ in $WX_2$. Dashed thin double arrows denote a diagonal energy shift (I, II, III) of the valley excitons, while the solid thick double arrows denote the off-diagonal coupling (IV) between different valley configurations.

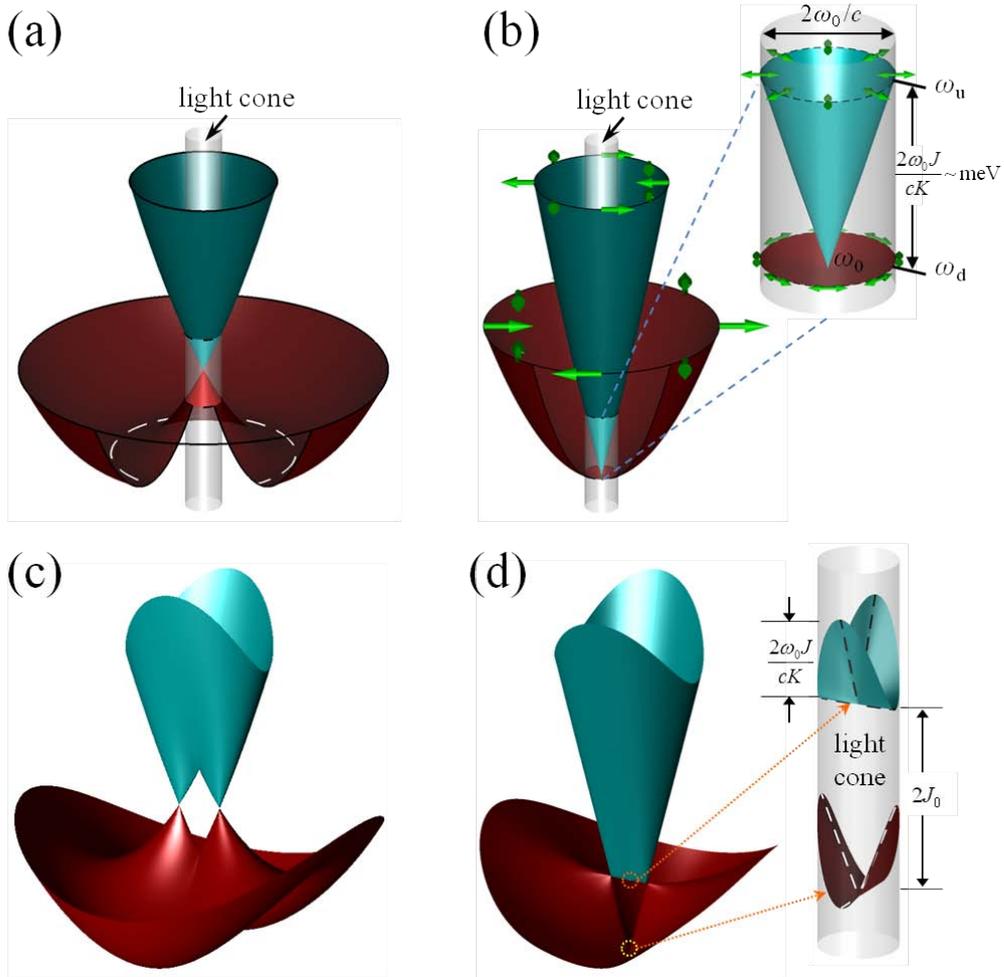

Fig. 4 | (a) Valley-orbit splitting of bright $X_0$ by the intervalley electron-hole exchange $J_{\mathbf{k}}^{\text{inter}}$ in the presence of the C3 rotational symmetry. (b) Valley-orbit coupled bright $X_0$ dispersions when both the inter-valley electron-hole exchange $J_{\mathbf{k}}^{\text{inter}}$ and the intra-valley electron-hole exchange $J_{\mathbf{k}}^{\text{intra}}$ are taking into account. The latter is a valley-independent energy shift that renormalize the dispersion of both the upper and lower branches. The single headed green arrows denote the excitonic valley pseudospin. The inset is the dispersion within the light cone, and the double headed arrows denote the linear polarization of the photon emission. (c) and (d): bright X0 dispersions in presence of a tensile strain (breaking C3 rotational symmetry), without (c) and with (d) the intra-valley exchange correction $J_{\mathbf{k}}^{\text{intra}}$, respectively. Part (a) and (c) are partly adapted from Ref. [32]. Copyright 2014, Nature Publishing Group.

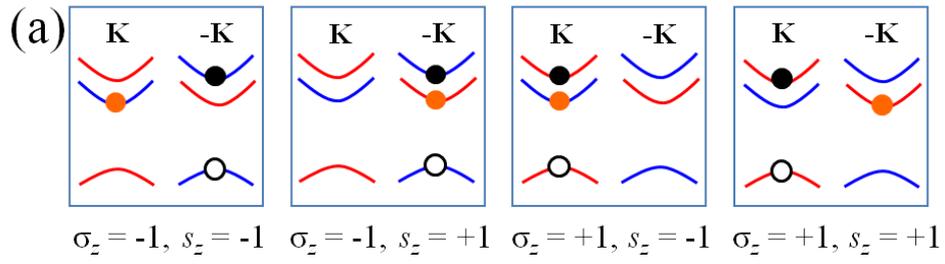

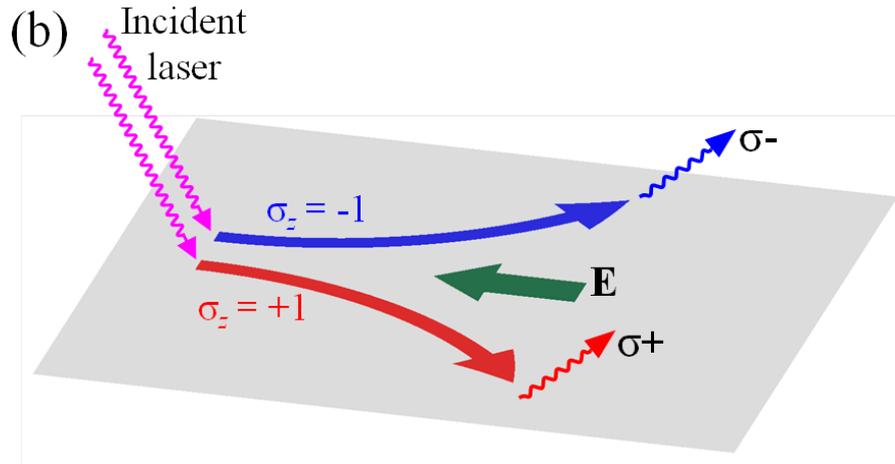

Fig. 5 | (a) The four valley configurations of bright X- characterized by the valley pseudospin $\hat{\sigma}_z$ of the recombining electron-hole pair, and spin $\hat{s}_z$ of the excess electron. (b) The valley Hall effect of the bright X-. Under an in-plane electric field **E**, X- with different $\hat{\sigma}_z$ will move to opposite transverse directions, and therefore the trion luminescence can have a spatially dependent polarization pattern.